\documentclass[11pt]{article}
\usepackage{epsfig}

\def\line#1{\hbox to \textwidth{#1}}
\catcode`\@=11

\def\thebibliography#1{\section*{REFERENCES}\list{\arabic{enumi}.}
  {\settowidth\labelwidth{#1.}\leftmargin=1.67em
   \labelsep\leftmargin \advance\labelsep-\labelwidth
   \itemsep\z@ \parsep\z@
   \usecounter{enumi}}\def\makelabel##1{\rlap{##1}\hss}%
   \def\newblock{\hskip 0.11em plus 0.33em minus -0.07em}
   \sloppy \clubpenalty=4000 \widowpenalty=4000 \sfcode`\.=1000\relax}

\def\@cite#1#2{$[{{#1\if@tempswa , #2\fi}}]$}

\newcount\@tempcntc
\def\@citex[#1]#2{\if@filesw\immediate\write\@auxout{\string\citation{#2}}\fi
  \@tempcnta\z@\@tempcntb\m@ne\def\@citea{}\@cite{%
        \@ordonner{#2}%
        \@for\@citeb:=#2\do%
    {\@ifundefined{b@\@citeb}%
        {\@citeo\@tempcntb\m@ne\@citea%
                \def\@citea{,\penalty\@m\ }{\bf ?}\@warning%
                {Citation `\@citeb' on page \thepage \space undefined}}%
        {\setbox\z@\hbox{\global\@tempcntc0\csname b@\@citeb\endcsname\relax}
     \ifnum\@tempcntc=\z@ \@citeo\@tempcntb\m@ne%
       \@citea\def\@citea{,\penalty\@m}%
       \hbox{\csname b@\@citeb\endcsname}%
     \else%
      \advance\@tempcntb\@ne%
      \ifnum\@tempcntb=\@tempcntc%
      \else\advance\@tempcntb\m@ne\@citeo%
      \@tempcnta\@tempcntc\@tempcntb\@tempcntc\fi\fi}}\@citeo}{#1}}%

\def\@citeo{\ifnum\@tempcnta>\@tempcntb\else\@citea
  \def\@citea{,\penalty\@m}%
  \ifnum\@tempcnta=\@tempcntb\the\@tempcnta\else
   {\advance\@tempcnta\@ne\ifnum\@tempcnta=\@tempcntb \else
\def\@citea{-}\fi
    \advance\@tempcnta\m@ne\the\@tempcnta\@citea\the\@tempcntb}\fi\fi}

\def\@toto{}
\newif\if@ordre 
\newcount\c@current
\newcount\c@last

\def\@ordonner#1{\global\c@last\m@ne%
                \global\@ordretrue%
                \@for\@toto:=#1\do%
                        {\@ifundefined{b@\@toto}%
                        {}%
                        {\c@current\csname b@\@toto\endcsname\relax%
                        \ifnum\the\c@current<\the\c@last\relax%
                                {\global\@ordrefalse}\fi%
                        \global\c@last\the\c@current%
                        }%
                        }%
                \if@ordre{}\else{\typeout{}%
                        \typeout{Warning: the references are not %
                         in increasing order\on@line:}%
                        \@for\@toto:=#1\do%
                        {\@ifundefined{b@\@toto}%
                        {}%
                        \typeout{\@toto:\space \@nameuse{b@\@toto}}%
                        }\typeout{}}\fi%
                }%

\catcode`\@=12

\catcode`\@=11

\newcount\c@subequation

\def\eqnarray{
\def\@eqnnum{{\reset@font\rm%
(\theequation-{\alph{subequation}})}}
\global\c@subequation=1\relax
\stepcounter{equation}\let\@currentlabel\theequation
\global\@eqnswtrue\m@th
\global\@eqcnt\z@\tabskip\@centering\let\\\@eqncr
$$\halign to\displaywidth\bgroup\@eqnsel\hskip\@centering
  $\displaystyle\tabskip\z@{##}$&\global\@eqcnt\@ne
  \hskip 2\arraycolsep \hfil${##}$\hfil
  &\global\@eqcnt\tw@ \hskip 2\arraycolsep $\displaystyle\tabskip\z@{##}$\hfil
   \tabskip\@centering&\llap{##}\tabskip\z@\cr}

\def\@@eqncr{\let\@tempa\relax
    \ifcase\@eqcnt \def\@tempa{& & &}\or \def\@tempa{& &}%
      \else \def\@tempa{&}\fi
     \@tempa \if@eqnsw\@eqnnum\global\advance\c@subequation by 1\relax
                        \fi
     \global\@eqnswtrue\global\@eqcnt\z@\cr}

\def\endeqnarray{\@@eqncr\egroup
      \global\advance\c@equation\m@ne$$\global\@ignoretrue
        \stepcounter{equation}
        \def\@eqnnum{{\reset@font\rm (\theequation)}}}

\def\Eqnarray{
\def\@eqnnum{{\reset@font\rm (\theequation)}}
\global\c@subequation=1\relax
\stepcounter{equation}\let\@currentlabel\theequation
\global\@eqnswtrue\m@th
\global\@eqcnt\z@\tabskip\@centering\let\\\@eqncr
$$\halign to\displaywidth\bgroup\@eqnsel\hskip\@centering
  $\displaystyle\tabskip\z@{##}$&\global\@eqcnt\@ne
  \hskip 2\arraycolsep \hfil${##}$\hfil
  &\global\@eqcnt\tw@ \hskip 2\arraycolsep $\displaystyle\tabskip\z@{##}$\hfil
   \tabskip\@centering&\llap{##}\tabskip\z@\cr}

\def\endEqnarray{\@@eqncr\egroup
      \global\advance\c@equation\m@ne$$\global\@ignoretrue
        \stepcounter{equation}
        \def\@eqnnum{{\reset@font\rm (\theequation)}}}

\catcode`\@=12

\font\tenmsa=msam10
\font\sevenmsa=msam7
\font\fivemsa=msam5
\font\tenmsb=msbm10
\font\sevenmsb=msbm7
\font\fivemsb=msbm5
\newfam\msafam
\newfam\msbfam
\textfont\msafam=\tenmsa  \scriptfont\msafam=\sevenmsa
\scriptscriptfont\msafam=\fivemsa
\textfont\msbfam=\tenmsb  \scriptfont\msbfam=\sevenmsb
\scriptscriptfont\msbfam=\fivemsb

\global\mathchardef\lesssim "142E

\newcommand{\slI}{\raise.15ex\hbox{$/$}\kern-.53em\hbox{$I$}}
\newcommand{\slL}{\raise.15ex\hbox{$/$}\kern-.53em\hbox{$L$}}
\newcommand{\slP}{\raise.15ex\hbox{$/$}\kern-.53em\hbox{$P$}}
\newcommand{\slR}{\raise.15ex\hbox{$/$}\kern-.53em\hbox{$R$}}
\newcommand{\slQ}{\raise.15ex\hbox{$/$}\kern-.53em\hbox{$Q$}}
\newcommand{\slK}{\raise.15ex\hbox{$/$}\kern-.53em\hbox{$K$}}
\newcommand{\slSigma}{\raise.15ex\hbox{$/$}\kern-.53em\hbox{$\Sigma$}}
\newcommand{\slcalP}{\raise.15ex\hbox{$/$}\kern-.63em\hbox{$\cal P$}}

\newcommand{\be}{\begin{equation}}
\newcommand{\ee}{\end{equation}}     
\newcommand{\bea}{\begin{eqnarray}}
\newcommand{\ena}{\end{eqnarray}}

\def\build#1\over#2{\mathrel{\mathop{\kern 0pt#1}\limits_{#2}}}

\font\tenimbf=cmmib10 at 12pt
\font\sevenimbf=cmmib10 at 7pt
\font\fiveimbf=cmmib10 at 5pt
\newfam\imbf
\textfont\imbf=\tenimbf
\scriptfont\imbf=\sevenimbf
\scriptscriptfont\imbf=\fiveimbf
\def\imb{\fam\imbf\tenimbf}

\begin{document}
\begin{titlepage}
\title{\begin{center}
\end{center}
\bf{Photon and lepton pair production in a quark-gluon plasma{\footnote {Based on talks given in July 2000 at QCD 00, Montpellier, France and at ICHEP2000, Osaka, Japan}}}}

\author{P.~Aurenche}
\maketitle

\begin{center}
\item Laboratoire de Physique Th\'eorique LAPTH{\footnote {UMR 5108 du CNRS, 
associ\'ee \`a l'Universit\'e de Savoie}},\\ 
BP110, F-74941, Annecy le Vieux Cedex, France
\end{center}

\vskip 3cm

\begin{abstract}

We discuss the production of real or virtual photons in a quark-gluon plasma.

\end{abstract}
\vfill
\centerline{\hfill LAPTH-Conf-810/2000}
\thispagestyle{empty}
\end{titlepage}


\section{Introduction}

It has long been thought that electro-magnetic probes {\em i.e.} real or
virtual photons would provide a way to detect the formation of a quark-gluon
plasma in ultra-relativistic heavy ion collisions. The energy distribution of
the photons would allow to measure the temperature of the plasma provided the 
rate of production in the plasma exceeds that of various backgrounds. It is
expected that this will occur in a small window in the GeV range for the energy
of the photon. At lower values of the energy the rate is dominated by various
hadron decay processes while at higher values the usual hard processes (those
occurring in the very early stage of the collision before the plasma is
formed), calculable by standard perturbative QCD methods, would dominate. In
contrast to hadronic observables (or heavy quarkonia) which are sensitive to
the late evolution of the plasma as well as to the re-hadronisation phase, the
photons in the GeV range are  produced soon after the plasma is formed and then
they escape the plasma without further interaction. 

We assume the plasma in thermal equilibrium (temperature T) with vanishing
chemical potential. The rate of production, per unit time and volume, of a real
photon of momentum $Q=(q_o,\imb q)$ is
\begin{equation}
  {{dN}\over{dtd{\imb x}}}=-
  {{d{\imb q}}\over{(2\pi)^3 2q_o}}\;2n_{_{B}}(q_o)\,
  {\rm Im}\,\Pi^{^{R}}{}_\mu{}^\mu(q_o,{\imb q})\; ,
  \label{realphot}
\end{equation}
while for a lepton pair of mass $\sqrt {Q^2}$ it is
\begin{equation}
  {{dN}\over{dtd{\imb x}}}=-
  {{dq_od{\imb q}}\over{12\pi^4}}\;
  {\alpha\over{Q^2}}\,n_{_{B}}(q_o)\,
  {\rm Im}\,\Pi^{^{R}}{}_\mu{}^\mu(q_o,{\imb q})\; ,
  \label{virtualphot}
\end{equation}
where $\Pi^{^{R}}{}_\mu{}^\mu(q_o,{\imb q})$ is the retarded photon
polarisation tensor. The pre-factor $n_{_{B}}(q_o)$ provides the expected
exponential damping $\exp(-q_o/T)$  when $q_o \gg T$. 
This report is devoted to the study of $\Pi^{^{R}}$ which contains the strong
interaction dynamics of quarks and gluons in the plasma. The theoretical
framework is that of the effective theory with re-summed hard thermal loops
(HTL)~\cite{BraatP1}.

We briefly review the status of  ${\rm Im}\,\Pi^{^{R}}$ calculated up to the two-loop approximation. Some phenomenological consequences are mentioned.
Then we turn to a discussion of higher loop corrections.

\section{The two-loop approximation}

Following the HTL approach~\cite{BraatP1} one distinguishes two scales: the
``hard" scale, typically of order $T$ or larger (the energy of quarks and
gluons in the plasma) and the ``soft" scale of order $g T$ where $g$, the
strong coupling, is assumed to be small. Collective effects in the plasma 
modify the physics at scale $g T$ {\em i.e.} over long distances of ${\cal
O}(1/gT)$. These effects lead to a modification of the propagators and vertices
of the theory  and one is led to introduce effective (re-summed) propagators
and vertices. This is easily illustrated with the example of the fermion
propagator, $S(P)$, which in the ``bare" theory is simply $1/p$ (we neglect
spin complications and make only a dimensional analysis). The thermal
contribution to the one loop correction $\Sigma(p)$ is found to be 
$\Sigma(p)\sim g^2 T^2 /p$ which is of the same order as the inverse propagator
when $p$ is of order $gT$. The re-summed propagator $^*S(P) = 1/ (p
-\Sigma(p))$ is then deeply modified for momenta of ${\cal O} (gT)$ whereas the
thermal corrections appear essentially as higher order effects for hard
momenta. Likewise, the gluon propagator and vertices are modified by hard
thermal loops when the external momenta are soft~\cite{BraatP1}. One can
construct an effective Lagrangian~\cite{BraatP4} in terms of effective
propagators and vertices and calculate observables in perturbation theory.

\begin{figure}
\epsfxsize120pt
\begin{center}
\epsfbox{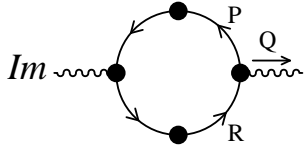}
\end{center}
\caption{\em One-loop contribution.}
\label{fig:1loop}
\end{figure}

In the one-loop approximation, the photon production rate is given by the diagram shown in fig.~\ref{fig:1loop} where the symbol $\bullet$ means that effective propagators and vertices are used. The result has been known for some time and can be expressed, in simplified notation, as~\cite{KapusLS1,AltheR1} 
\begin{equation} 
{\rm Im}\,\Pi^{^{R}} (q_o,{\imb q}) \sim e^2 g^2 T^2
\Big(\ln ({q_0 T\over m_q^2}) + C({Q^2 \over m_q^2}) \Big) 
\label{eq:1loop}
\end{equation} 
where $m_q^2 \sim g^2 T^2$ is related to the thermal mass of the
quark. One notes the presence of a ``large" logarithmic term $\ln(1/g)$
dominating over a ``constant term" $C(Q^2 / m_q^2)$.

\begin{figure}[hbt]
\epsfxsize240pt
\begin{center}
\epsfbox{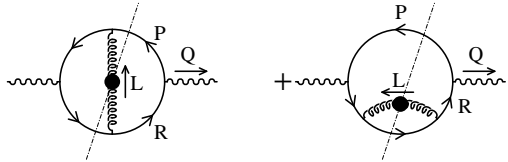}
\end{center}
\caption{\em The dominant two-loop contributions.}
\label{fig:2loop}
\end{figure}

\begin{figure}
\epsfxsize240pt
\begin{center}
\hspace{-15pt}
\epsfbox{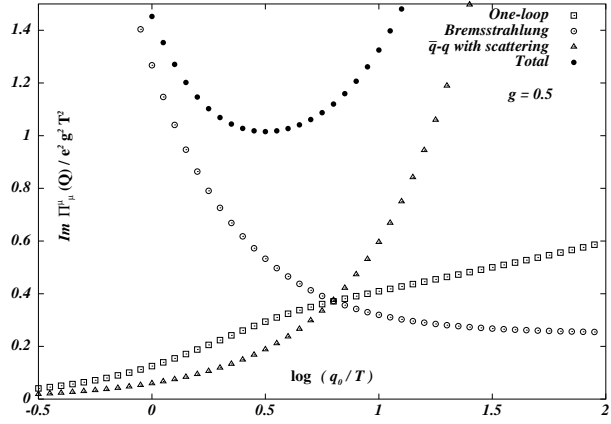}
\end{center}
\caption{\em Comparison of various contributions to ${\rm
  Im}\,\Pi{}^\mu_\mu(Q)$ for a hard real photon. The comparison is made for
  $N=3$ colors and $N_{_{F}}=2$ flavors.}
\label{fig:compar}
\end{figure}

The two-loop diagrams are displayed in fig.~\ref{fig:2loop}. In principle,
there are more diagrams in the effective theory but only those leading to the
dominant contribution are shown. All propagators and vertices should be
effective but since the largest contribution arises from hard fermions it is
enough, following the HTL strategy, to keep bare fermion propagators and
\begin{figure}
\epsfxsize280pt
\begin{center}
\epsfbox{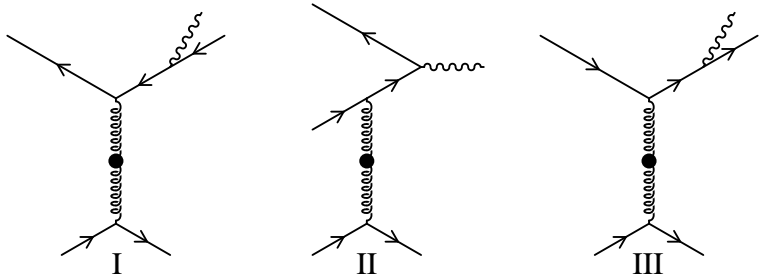}
\end{center}
\vskip -15pt
\caption{\em Physical processes included in the diagrams of
      Fig.~\ref{fig:2loop}, in the region $L^2<0$.
      I:  bremsstrahlung with an antiquark. II: $q\bar{q}$ annihilation
      with scattering. III: bremsstrahlung with a quark.}
\label{fig:processes}
\end{figure}
vertices as indicated{\footnote {Note that for consistency of our approach,
based on an expansion in terms of effective quantities, we keep the thermally
generated mass in the hard limit of the effective propagator.}}.  Only the
gluon line needs to be effective since soft momentum $L$ through the gluon line
dominates the integrals. To evaluate  these diagrams it is convenient to
distinguish between the contribution arising from a time-like gluon ($L^2 >0$)
and a space like gluon ($L^2 < 0$). The first type leads to a contribution
similar to eq.~(\ref{eq:1loop}) and requires some care as counter-terms (not
shown) eliminate the parts of the two-loop diagrams already contained in the
one-loop diagrams~\cite{AurenGKZ2}. We concentrate on the second case  which in
terms of physical processes corresponds to bremsstrahlung production  of a
photon or production in a quark-antiquark annihilation process where one of the
quark is put off-shell by scattering in the plasma (see
fig.~\ref{fig:processes}). The result for hard photons is~\cite{AurenGKZ1}
\begin{eqnarray}
{\rm Im}\,\Pi^{^{R}} (q_o,{\imb q})\Big|_{\rm brems} \sim e^2\ g^2 T^2 \\
{\rm Im}\,\Pi^{^{R}} (q_o,{\imb q})\Big|_{\rm annil} \sim e^2\ g^2 T q_0
\label{eq:2loop}
\end{eqnarray}
The reason why these two-loop contributions have the same order as the one-loop
one is due to the presence of strong collinear singularities. To calculate
${\rm Im}\,\Pi^{^{R}}$ one has to cut the propagators as indicated by the
dash-dotted lines in fig.~\ref{fig:2loop}. In the integration over the loop
hard momentum $P$ (with $P^2$, $(R+L)^2$ on shell) the denominators $R^2$ and
$(P+L)^2$ of the un-cut fermion propagators simultaneously almost vanish when
$\imb p$ is parallel to $\imb q$ {\em i.e.} in the collinear configuration.
This leads to an enhancement factor of type $T^2/M^2_{\rm eff}$ where the
cut-off $M^2_{\rm eff}=  m_q^2 + p (p+q_0) Q^2/q^2_0$ emerges from the
calculation. For the kinematic range of concern to us here, $M^2_{\rm eff} \sim
g^2 T^2$ so that the two-loop diagram is enhanced by a factor $1/g^2$ which
compensates the  $g^2$ factor associated to the coupling of the gluon to the
quarks. An interesting result of the calculation is the importance of process
II of fig.~\ref{fig:processes} which grows with the energy of the photon and
dominates over the other contributions when $q_0/T \gg 1$ as shown in
fig.~\ref{fig:compar}. 
Phenomenological applications of these results have been carried out and
the two-loop processes have been included in hydrodynamic evolution codes to
predict the rate of real photon production at RHIC or LHC~\cite{MustaT}. It is
found that the two-loop processes (especially the annihilation with
scattering) lead to an increase by an order of magnitude compared to the
one-loop processes. This may even have consequences for heavy ion collisions at
SPS energies~\cite{Sriva}. Several effects may reduce these over-optimistic
predictions: lack of chemical equilibrium and more importantly higher order
corrections as discussed next. 

\section{Higher order corrections}

\begin{figure}
\epsfxsize200pt
\begin{center}
\epsfbox{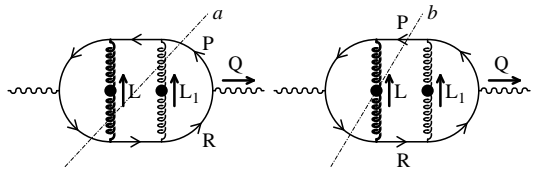}
\end{center}
\caption{\em A ladder diagram.}
\label{fig:ladder}
\end{figure}

Since the one-loop and two-loop results are of the same order it is reasonable
to worry about the convergence of the perturbative expansion in the effective
theory! The enhancement mechanism operative at two-loop could also be at work at
the multi-loop level especially in ladder diagrams, an example of which is shown
in fig.~\ref{fig:ladder}: indeed many ``small" fermion denominators appear in
such diagrams which can produce a pile-up of collinear singularities.
A recent study of the three-loop ladder diagram shows that~\cite{AurenGZ1}
\begin{equation}
{\rm Im}\,\Pi^{^{R}} \Big|_{\rm 3-loop} \sim
{\rm Im}\,\Pi^{^{R}} \Big|_{\rm 2-loop} \times {g^2 T \over l_{min}}
\label{eq:3loop}
\end{equation}
where $l_{min}$ is the largest of the cut-offs:\\
-- $l^{(1)}_{min} = M^2_{\rm eff} q_0 / p_0 r_0$, which is the collinear
cut-off encountered above: it depends on the thermal quark mass and momentum
($p_0 \sim r_0 \sim T$) as well as on the external variables;\\
-- $l^{(2)}_{min} = m_D \sim g T$, the Debye mass if the added gluon is longitudinal, or $l^{(2)}_{min} = m_{mag} \sim g^2 T$ if it is transverse.\\
For the kinematic configuration of interest, in the case of an extra
longitudinal gluon one can check that $m_D \gg l^{(1)}_{min}$ and the Debye
mass acts as a cut-off with the result that the three-loop contribution is 
suppressed by a factor $g$ compared to the two-loop. On the contrary, for a
transverse gluon, both regulators are of order $g^2 T$ (as long as $Q^2/q_0^2 <
g^2 $) and the three-loop diagram is of the same order as the two-loop one. One
is therefore in a non-perturbative regime. The problem is similar to the
magnetic mass problem pointed out by Linde in the perturbative calculation of
the free energy~\cite{Linde}, except that here it appears at leading order. 
\begin{figure}
\epsfxsize240pt
\vspace{-5pt}
\begin{center}
\hspace{-7pt}
\epsfbox{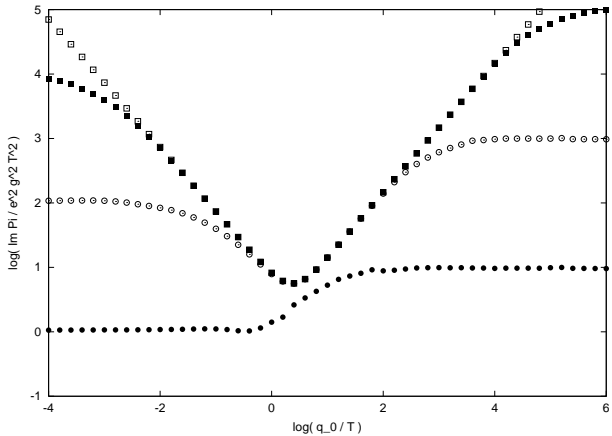}
\end{center}
\caption{\em Effect of the width $\Gamma$ on the two-loop contributions. Each curve
  corresponds to a different values of $\Gamma$. From top to bottom, the ratio
  $\Gamma T / m^2_q$ takes the values 10$^{-6}$, 10$^{-4}$, 10$^{-2}$ and 1.}
\label{fig:damping}
\end{figure}
\begin{figure}
\epsfxsize210pt
\begin{center}
\epsfbox{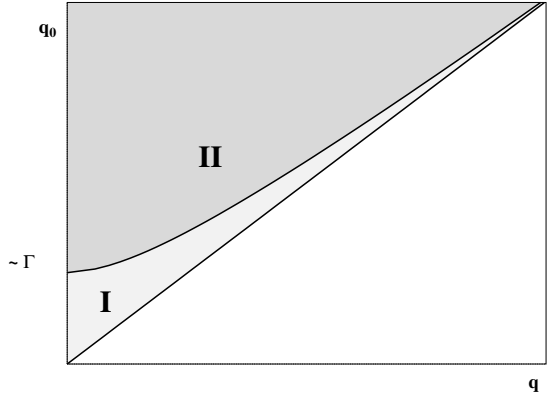}
\end{center}
\caption{\em Boundary obtained from the condition  $\widehat{\Gamma}=1$. In region
  I, the width is the dominant regulator of collinear singularities. In region
  II, the width is only a sub-dominant correction.}
\label{fig:2}
\end{figure}

Another effect which can modify the collinear enhancement mechanism is related
to the fermion damping rate. Indeed, including the damping rate on the fermion
lines, will shift the pole of the propagators away from the real axis: this
affects the enhancement mechanism based on the near-vanishing of the
denominators. Ignoring the requirement of gauge invariance and concentrating
only on the  mathematical effect of shifting the poles to the complex plane one
can do again the two-loop calculation with fermion propagators including the
damping rate $\Gamma \sim g^2 T \ln(1/g)$. The result is intuitively simple as
a regulator of the form~\cite{AurenGZ2}
\begin{equation}
{\cal M}^2_{\rm eff} = M^2_{\rm eff} + 4 i \Gamma {p_0 (p_0+q_0) \over q_0}
\label{eq:damping}
\end{equation}
comes out, with $M^2_{\rm eff}$ defined above.  The effect of $\Gamma$ on
${\rm Im}\,\Pi^{^{R}} (q_o,{\imb q})\big|_{\rm 2-loop}$ is shown on
fig.~\ref{fig:damping} for the case of a real photon (${Q^2=0}$). The region
$q_o/T < 1$ is dominated by bremsstrahlung emission while the region $q_o/T >
1$  receives a contribution mainly from the annihilation with scattering
process (see fig.~\ref{fig:compar}). The top curve is the result obtained with
a vanishingly small width. One notes the change in the $q_o$ behaviour of ${\rm
Im}\,\Pi^{^{R}} (q_o,{\imb q})$ as $\Gamma$ increases: this is due to the
different $q_o$ dependences of the real and imaginary parts of ${\cal
M}^2_{\rm eff}$. For virtual photon production, one notes that the quantity 
$\big|{\rm Re}\ {\cal M}^2_{\rm eff}\big|$ increases with $Q^2$ at fixed $q_o$
so that the ratio ${\widehat \Gamma} = {\rm Im}\ {\cal M}^2_{\rm eff} / {\rm
Re}\ {\cal M}^2_{\rm eff}$, which controls the relative importance of the
width, decreases. For $Q^2$ large enough the effect of $\Gamma$ will become
negligible and the two-loop calculation should be adequate. This is illustrated
in fig.~\ref{fig:2}.

Equation~(\ref{eq:damping}) lends itself to a simple interpretation. It
can be written as
\begin{equation}
{\cal M}^2_{\rm eff} = 2 {p_0 (p_0+q_0) \over q_o} 
\big(1/ \lambda_{\rm for} + i/ \lambda_{\rm mean} \Big)
\label{eq:intuit}
\end{equation}
where $\lambda_{\rm mean}= 1/\Gamma$ is the mean free path of the quark in the
plasma and $\lambda_{\rm for}=  2 p_0(p_0+q_0)/M^2_{\rm eff} q_0 $ can be shown
to be  the formation length of the photon. Then, if $\lambda_{\rm mean} \gg
\lambda_{\rm for}$ the effect of the damping rate can be ignored and the
corresponding higher order diagrams are suppressed. In the opposite case,
re-scattering in the plasma modifies the two-loop result. This is equivalent to
say that the Landau-Pomeranchuk-Migdal (LPM) effect~\cite{LandaP2} has to be
taken into account in the calculation. Two interesting features emerge from the
above discussion: 1) the LPM effect not only modifies the production of
bremsstrahlung photon but also that of very hard photons emitted in the
``annihilation with scattering" process as illustrated in
fig.~\ref{fig:damping}; 2) if the virtuality ${Q^2}$ of the hard lepton pair is
large enough then one falls in the domain $\lambda_{\rm mean} \gg \lambda_{\rm
for}$ and the perturbative calculation at two-loop is sufficient.  

The problems discussed above are an illustration of a more general situation
concerning thermal Green's function with external momenta close to the
light-cone~\cite{Gelis}.

The production mechanism of hard photons in the plasma is very complex. New
processes appear at two-loop which considerably increase the rate of photon
production calculated at one-loop. However, for real or small mass virtual
photons the higher loop diagrams become important and the rate  turns out to be
non-perturbative. Taking into account higher order effects to obtain a
quantitative estimate remains to be done.

\section*{Acknowledgments}
I thank F. Gelis, R. Kobes and H. Zaraket for a fruitful collaboration on the
work discussed above.

\end{document}